\documentclass[apj]{emulateapj}
\usepackage{pstricks,amsmath}
\usepackage{apjfonts}

\def\1e{\mbox{1E\,0657--56}}
\def\rtr{r_{\rm tr}}
\def\vs{v_{\rm s}}
\def\kms        {km$\,$s$^{-1}$}
\def\hseventy   {$H_0$=70~km$\;$s$^{-1}\,$Mpc$^{-1}$}
\def\cmg        {cm$^{2}\,$g$^{-1}$}
\def\chandra    {{\em Chandra}\/}
\def\bi{\bfseries\itshape}
\def\gcmcube    {g$\,$cm$^{-3}$}
\def\gcmsq      {g$\,$cm$^{-2}$}
\def\lax{\lesssim}

\def\msun       {$M_{\odot}$}
\def\deg        {$^{\circ}$}

\begin{document}

\submitted{ApJ in press; ~astro-ph/0309303 v2}

\lefthead{DARK MATTER SELF-INTERACTION CROSS-SECTION}
\righthead{MARKEVITCH ET AL.}

\title{DIRECT CONSTRAINTS ON THE DARK MATTER SELF-INTERACTION CROSS-SECTION
FROM THE MERGING GALAXY CLUSTER 1E\,0657--56}

\author{M.~Markevitch\altaffilmark{1}, 
A.~H.~Gonzalez\altaffilmark{2},
D.~Clowe\altaffilmark{3,4}, 
A.~Vikhlinin\altaffilmark{1,5}, 
W.~Forman\altaffilmark{1}, 
C.~Jones\altaffilmark{1}, 
S.~Murray\altaffilmark{1},
W.~Tucker\altaffilmark{1,6}}

\altaffiltext{1}{Harvard-Smithsonian Center for Astrophysics, 60 Garden St.,
Cambridge, MA 02138; maxim@head.cfa.harvard.edu}

\altaffiltext{2}{Department of Astronomy, University of Florida}

\altaffiltext{3}{Institut f\"ur Astrophysik und Extraterrestrische Forschung
  der Universit\"at Bonn, Germany}

\altaffiltext{4}{Stewart Observatory, University of Arizona}

\altaffiltext{5}{IKI, Moscow, Russia} 

\altaffiltext{6}{University of California at San Diego} 

\setcounter{footnote}{6}

\begin{abstract}

We compare new maps of the hot gas, dark matter, and galaxies for \1e, a
cluster with a rare, high-velocity merger occurring nearly in the plane of
the sky. The X-ray observations reveal a bullet-like gas subcluster just
exiting the collision site.  A prominent bow shock gives an estimate of the
subcluster velocity, 4500 \kms, which lies mostly in the plane of the
sky. The optical image shows that the gas lags behind the subcluster
galaxies.  The weak-lensing mass map reveals a dark matter clump lying ahead
of the collisional gas bullet, but coincident with the effectively
collisionless galaxies.  From these observations, one can directly estimate
the cross-section of the dark matter self-interaction.  That the dark matter
is not fluid-like is seen directly in the X-ray -- lensing mass overlay;
more quantitative limits can be derived from three simple independent
arguments.  The most sensitive constraint, $\sigma/m<1$ \cmg, comes from the
consistency of the subcluster mass-to-light ratio with the main cluster (and
universal) value, which rules out a significant mass loss due to dark matter
particle collisions.  This limit excludes most of the $0.5-5$ \cmg\ interval
proposed to explain the flat mass profiles in galaxies.  Our result is only
an order-of-magnitude estimate which involves a number of simplifying, but
always conservative, assumptions; stronger constraints may be derived using
hydrodynamic simulations of this cluster.

\end{abstract}


\section{INTRODUCTION}
\label{sec:intro}

\1e, one of the hottest and most X-ray luminous galaxy clusters known, was
discovered by Tucker et al.\ (1995).  It was first observed by \chandra\ in
October 2000 for 24 ks. That observation revealed a bullet-like, relatively
cool subcluster just exiting the core of the main cluster, with a prominent
bow shock (Markevitch et al.\ 2002, hereafter M02).  A comparison of the
X-ray and optical images revealed a galaxy subcluster just ahead of the gas
``bullet'', which led M02 to suggest that this unique system could be used
to determine whether dark matter is collisional or collisionless, if only
one could map the mass distribution in the subcluster.  Apart from the
obvious interest for the still unknown nature of dark matter, the
possibility of it having a nonzero self-interaction cross-section has
far-reaching astrophysical implications (Spergel \& Steinhardt 2000; for
more discussion see \S\ref{sec:disc} below).

\begin{figure*}[t]
\pspicture(0,14.9)(18.5,24.0)

\rput[tl]{0}(-0.5,24){\epsfxsize=10.0cm \epsfclipon
\epsffile{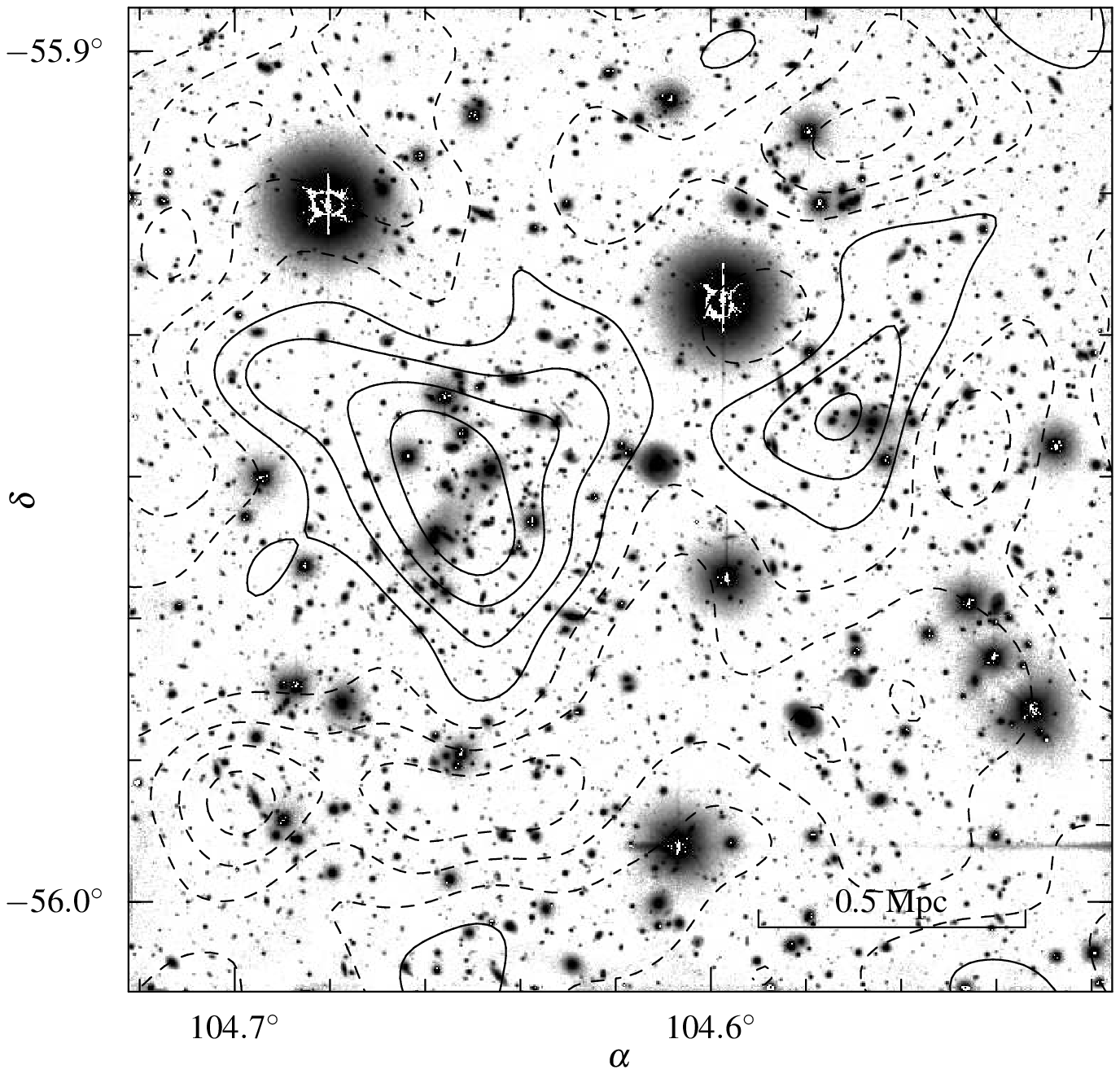}}

\rput[tl]{0}(8,24){\epsfxsize=10.0cm \epsfclipon
\epsffile{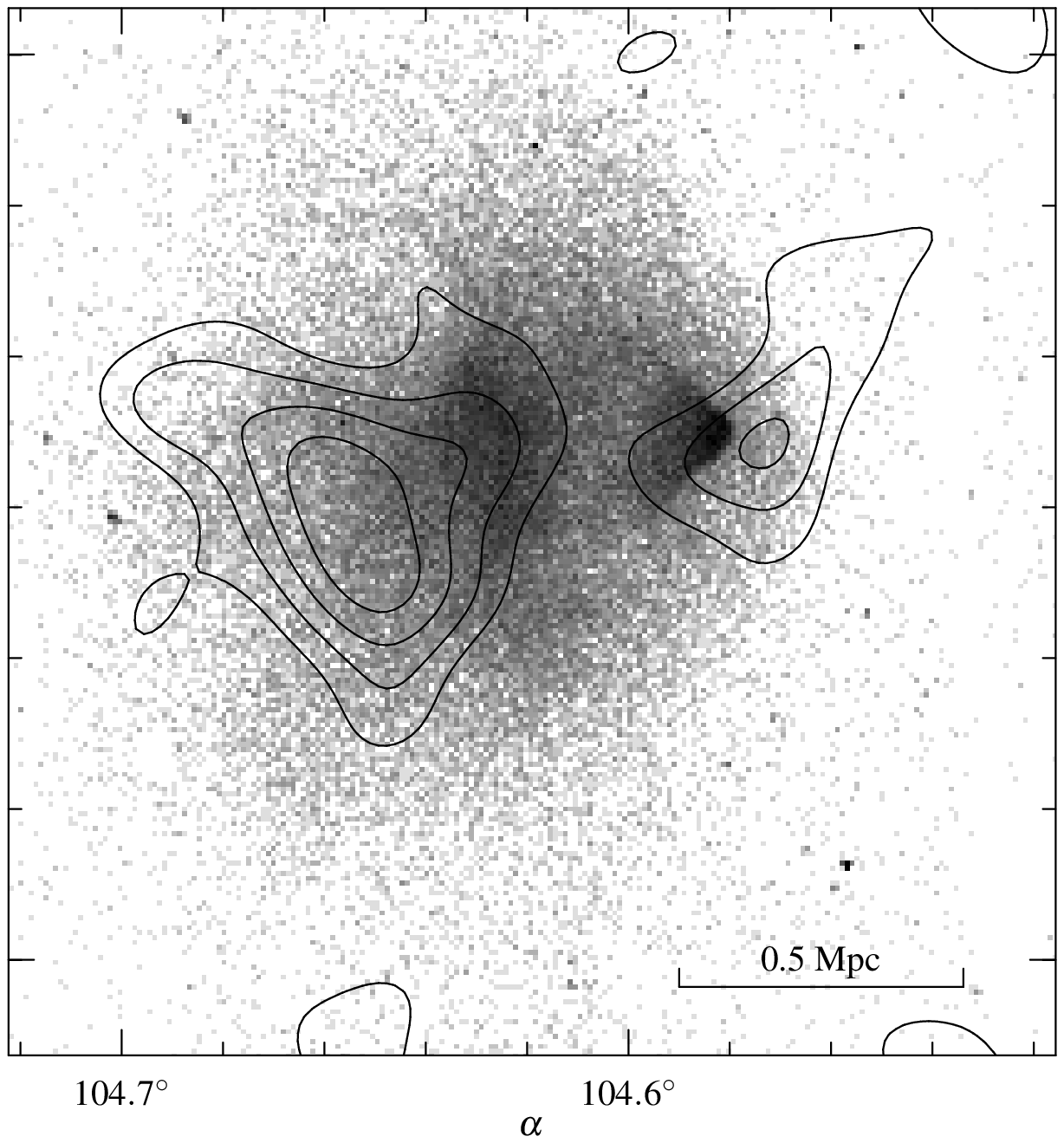}}

\rput[bl]{0}(1.7,23.0){\large\bi a}
\rput[bl]{0}(10.2,23.0){\large\bi b}

\endpspicture

\caption{({\em a}) Overlay of the weak lensing mass contours
  on the optical image of \1e. Dashed contours are negative (relative to an
  arbitrary zero level).  The subcluster's DM peak is coincident within
  uncertainties with the centroid of the galaxy concentration. ({\em b})
  Overlay of the mass contours on the X-ray image (only the upper 4 of those
  in panel {\em a}\/ are shown for clarity). The gas bullet lags behind the
  DM subcluster.}
\label{fig:img}
\end{figure*}

Just such a map of the dark matter distribution in \1e\ has recently been
obtained by Clowe, Gonzalez, \& Markevitch (2004, hereafter C04) from weak
lensing data. It reveals a dark matter clump coincident with the centroid of
the galaxies (Fig.\ \ref{fig:img}{\em a}).  C04 also derived $M/L$ ratios of
the main cluster and the subcluster and found them in agreement with each
other and with other clusters' values. In addition, \chandra\ re-observed
\1e\ for 70 ks in July 2002, from which a more accurate estimate of the
shock Mach number was derived using the gas density jump at the shock,
$M=3.2^{+0.8}_{-0.6}$ (all uncertainties 68\%), which corresponds to a shock
(and bullet subcluster) velocity of $\vs=4500^{+1100}_{-800}$ \kms\
(Markevitch et al., in prep., hereafter M04).  The new X-ray data also
further clarified the geometry of the merger.  

In this paper, we combine these new optical and X-ray data to constrain the
self-interaction cross-section of dark matter particles.  We use
$\Omega_0=0.3$, $\Omega_\Lambda=0.7$, \hseventy, for which $1''=4.42$ kpc at
the cluster redshift $z=0.296$.

\section{COLLISIONAL CROSS-SECTION ESTIMATES}

The dark matter collisional cross-section, $\sigma$, can be constrained from
the \1e\ data by at least three independent methods, using simple
calculations described in the sections below.  They are based on the
observed gas--dark matter offset, the high subcluster velocity, and the
subcluster survival.  First, we give the main assumptions that will go into
these calculations.

There are two estimates of the total masses of the subcluster and the main
cluster --- from the galaxy velocity dispersion (Barrena et al.\ 2002) and
from weak lensing (C04).  Given the disturbed state of this system, virial
or hydrostatic mass estimates (either from galaxy velocities or the gas
temperature) can be incorrect, and we chose to use the direct weak lensing
measurements from C04 even though their formal statistical accuracy is
poorer.  The main cluster's lensing signal can be fit by a King mass profile
$\rho=\rho_0 (1+r^2/r_c^2)^{-3/2}$ with best-fit parameters $\rho_0\simeq
2.6\times 10^{-25}$ \gcmcube\ and $r_c\simeq 210$ kpc (C04).  These two
parameters are degenerate so their individual error bars are not meaningful;
the quantity of interest to us is the central mass column density
(approximately proportional to $\rho_0 r_c$), which is measured with a
16\% accuracy.  This mass profile is very close to the Barrena et
al.\ NFW profile at all radii outside the core.  A King profile is
marginally preferred over an NFW profile (also acceptable statistically).

The projected mass excess created by the subcluster is detected in the
lensing data with a $3.0\sigma$ significance.  The subcluster mass signal is
detected to $r\simeq 150-200$ kpc from the mass (or galaxy) peak, beyond
which the subcluster may be tidally stripped (C04).  The lensing-derived
subcluster mass ($M\simeq 7\times 10^{13}$\msun) is significantly higher
than the Barrena et al.\ estimate, but the latter result was based on only 7
galaxies and the equilibrium assumption, thus could easily be biased.  The
current lensing data accuracy is not sufficient to derive an exact mass
distribution for the subcluster, but our estimates below are not
particularly sensitive to it, being mostly determined by its overall
projected mass.  For the sake of modeling, we will adopt a King profile with
$r_c=70$ kpc and $\rho_0=1.3\times 10^{-24}$ \gcmcube, truncated at
$\rtr=150$ kpc, which adequately describes the lensing data.

The subcluster is assumed to have passed (once) close to the center of the
main cluster.  This is supported by the X-ray image (Fig.\ \ref{fig:img}{\em
b}), the gas temperature map (M04) and their comparison with the radio halo
map (Govoni et al.\ 2004; Liang et al.\ 2000).  A cooler North-South bar in
the X-ray image between the two dark matter clumps appears to be an edge-on
pancake-like remnant of the merged main cluster's gas core and the
subcluster's outer atmosphere (stripped from what is now the gas bullet),
suggesting that the subcluster has passed straight through the densest
cluster region (M04).  The line-of-sight velocity of the subcluster relative
to the main cluster is about 600 \kms\ (Barrena et al.\ 2002); combined with
the X-ray-measured Mach number, it gives an angle of only $\sim 8$\deg\
between the direction of motion and the plane of the sky.  The sharpness of
the shock front also confirms that the subcluster is presently moving very
nearly in the plane of the sky (M02).  From all the above, it is reasonable
to assume that the subcluster has passed through the core of the main
cluster.

The accuracy of our qualitative cross-section estimates will be determined
by the validity of this and other assumptions (given below where relevant)
to a greater degree than by the measurement uncertainties, so below we will
omit the measurement error propagation for clarity.

\subsection{The gas --- dark matter offset}

The most remarkable feature in Fig.\ \ref{fig:img}{\em b}\/ is a $\sim 23''$
offset between the subcluster's DM centroid and the gas bullet, which is at
least $2\sigma$-significant (C04). C04 use this fact as a direct proof of
dark matter existence, as opposed to modified gravity hypotheses (Milgrom
1983 and later works) in which one would expect the lensing mass peak to be
associated with the gas --- the dominant visible mass component.  For our
purposes, this offset means that the scattering depth of the dark matter
subcluster w.r.t.\ collisions with the flow of dark matter particles cannot
be much greater than 1.  Otherwise the DM subcluster would behave as a clump
of fluid, experiencing stripping and drag deceleration, similar to that of
the gas bullet (assuming the same gas mass fraction in the main cluster and
the subcluster), and there would be no offset between the gas and dark
matter.  The subcluster's scattering depth is
\begin{equation}
\tau_{\rm s} = \frac{\sigma}{m}\, \Sigma_{\rm s},
\end{equation}
where $\sigma$ is the DM collision cross-section, $m$ is its particle mass,
and $\Sigma_{\rm s}$ is the DM mass surface density of the subcluster.  The
surface density averaged over the face of the subcluster within $r=r_{\rm
tr}$ is $\Sigma_{\rm s}\simeq 0.2$ \gcmsq.  Assuming spherical symmetry and
requiring that $\tau_{\rm s}<1$, we obtain
\begin{equation}
\frac{\sigma}{m}<5\;{\rm cm}^2\,{\rm g}^{-1}.
\end{equation}
The surface density toward the subcluster center is several times higher, so
by using an average we obtain a conservative upper limit.

Another remarkable feature in Fig.\ \ref{fig:img}{\em a}\/ is the
coincidence of the subcluster's DM and galaxy centroids within their
uncertainties (C04).  To avoid an easily made mistake, we should note here
that the subcluster galaxies, although effectively collisionless, do not
give us the position at which the subcluster would be in the absence of the
DM collisions.  The galaxies are gravitationally bound to the DM clump and
will be dragged back by it, should it experience any deceleration from the
DM collisions.  This can be taken into account, and a limit on the offset
between the DM and galaxy centroids can indeed be used to derive an
independent $\sigma/m$ constraint; however, it requires more accurate
centroid positions than currently available.

\subsection{The high velocity of the subcluster}
\label{sec:vel}

The observed velocity of the subcluster, $\vs=4500$ \kms, is in good
agreement with the expected free-fall velocity onto the main cluster.  For
our main cluster's mass profile, a small subcluster falling from a large
distance should acquire 4400 \kms\ at core passage, decelerating to 3500
\kms\ at the current 0.66 Mpc off-center distance of the subcluster.  Since
the cluster peculiar velocities are small, such an agreement strongly
suggests that the subcluster could not have lost much of its momentum to
drag forces.  Drag is created by putative DM particle collisions, as well as
the gravitational pull of the gas being stripped from the subcluster and of
the tidally-stripped outer subcluster mass, and by dynamical friction as the
moving subcluster disturbs the main cluster's matter distribution.  We will
conservatively disregard the latter three (all of which are relatively small
effects) and assume, for a qualitative estimate, that the loss of velocity
due to the DM collisions, compared to free fall, is less than 1000 \kms,
accumulated along the way through the main cluster.

\begin{figure}[b]
\pspicture(0,20.)(18.5,23.5)

\rput[tl]{0}(0,24){\epsfxsize=8.0cm \epsfclipon
\epsffile[85 333 550 583]{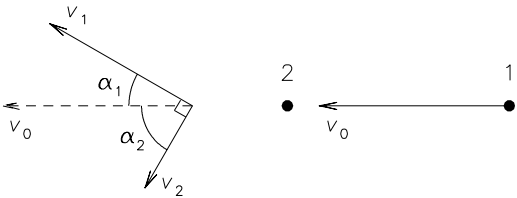}}
\endpspicture

\caption{Collision of two equal-mass particles in the subcluster reference
  frame.} 
\label{fig:diagram}
\end{figure}

We will now derive the drag force on the subcluster from the DM particle
collisions.  The subcluster will be decelerated by collisions when its
particles acquire a momentum component opposite to the subcluster's velocity
and then transfer all or part of it to the whole subcluster via
gravitational interactions.  We assume for simplicity that (a) dark matter
particles have no peculiar velocities, and (b) the subcluster's gravity is
felt by the particles, and vice versa, as long as the particle is within a
fiducial radius $r'=2\rtr$\/ of the subcluster (the result will not depend
qualitatively on $r'$ in the range $\rtr-\infty$).  We further assume that a
particle transfers all of its momentum to the subcluster if its
post-collision velocity is insufficient to escape beyond $r'$.
Faster-scattered particles retard the cluster only until they reach $r'$.
The critical velocity that a subcluster particle needs to reach $r'$,
mass-averaged (in the {\em rms}\/ sense) over the subcluster mass profile,
is $V\simeq 1900$ \kms.  Use of this average value lets us make a further
simplifying assumption that all collisions occur at the subcluster center.
We also conservatively disregard multiple scattering in the subcluster; the
adequacy of this assumption will be addressed below.

An elastic collision of two equal-mass particles proceeds as shown in Fig.\
\ref{fig:diagram}. In the subcluster's reference frame, particle 2 is at
rest and particle 1, in the incoming flow, collides with it with a velocity
\begin{equation}
v_0=\sqrt{\vs^2+V^2} \approx 4800 \;{\rm km\;s}^{-1},
\label{eq:v0}
\end{equation}
where $\vs$ is the subcluster's velocity and the relatively small increase
is because of the subcluster's gravitational pull within $r=r'$.  Particle 1
scatters at an angle $\alpha_1>0$ with a velocity $v_1$, while particle 2
acquires a velocity $v_2$ at an angle $\alpha_2>0$. From energy and momentum
conservation, $\alpha_1+\alpha_2=\pi/2$ and the velocities are
\begin{equation}
v_1 = v_0 \cos \alpha_1, ~~~
v_2 = v_0 \sin \alpha_1.
\label{eq:v}
\end{equation}

As a result of each collision, the subcluster acquires a net momentum along
the $v_0$ direction, $p=p_0+p_1+p_2$, where
\begin{equation}
p_0=m\,(\vs-v_0)
\label{eq:p0}
\end{equation}
comes from the infalling particle (which carries back exactly $-p_0$ if it
passes through the subcluster without scattering), and $p_i$\/ ($i=1,2$)
comes from the respective outbound particle.  If $v_i<V$ then
\begin{equation}
p_i=m\,v_i\,\cos\alpha_i
\label{eq:pall}
\end{equation}
(the particle transfers all its momentum to the subcluster), while for
$v_i>V$,
\begin{equation}
p_i=m \left[ v_i - (v_i^2-V^2)^{1/2} \right] \cos\alpha_i.
\label{eq:pesc}
\end{equation}
For our particular mass and velocity, at least one particle always escapes
beyond $r'$. If both escape, which occurs in collisions with $\alpha_{\rm
e}<\alpha_1<\frac{\pi}{2}-\alpha_{\rm e}$\; ($\alpha_{\rm e}\simeq 23^\circ$
for our parameters), the momentum loss by the subcluster as a function of
the scattering angle $\alpha_1$ is (combining eqs.\ \ref{eq:p0},
\ref{eq:pesc}, \ref{eq:v0}, and \ref{eq:v}):
\begin{equation}
\begin{split}
p_{\rm both}  = m \vs \bigg[ 1  & 
- \cos\alpha_1 
\left(\cos^2\alpha_1-\frac{V^2}{\vs^2}\sin^2\alpha_1\right)^{1/2} \\
& \left. 
- \sin\alpha_1
\left(\sin^2\alpha_1-\frac{V^2}{\vs^2}\cos^2\alpha_1\right)^{1/2}\right].
\label{eq:pboth}
\end{split}
\end{equation}
If only one particle escapes, which occurs when
$0<\alpha_1<\alpha_{\rm e}$\; or\; $\frac{\pi}{2}-\alpha_{\rm
e}<\alpha_1<\frac{\pi}{2}$, the net momentum loss is (combining eqs.\
\ref{eq:p0}, \ref{eq:pall}, \ref{eq:pesc}, \ref{eq:v0}, and \ref{eq:v}):
\begin{equation}
p_{\rm one}=m\vs \left[1- \cos\alpha_1 
\left(\cos^2\alpha_1-\frac{V^2}{\vs^2}\sin^2\alpha_1\right)^{1/2}\right].
\label{eq:pone}
\end{equation}

Now we should average the above momentum loss over all scattering angles.
For this, it is convenient to use the reference frame of the center of mass
of the colliding particles.  In this frame, the scattering is isotropic, as
long as the particles are ``slow'' in the sense that $m u r /\hbar \ll 1$,
where $r$\/ is the linear scale of the interaction and $u=v_0/2$\/ is the
collision velocity (Landau \& Lifshitz 1958, \S132).  For example, for rigid
spheres with radius $r$\/ (for which the scattering cross-section is
$\sigma=4\pi r^2$, Landau \& Lifshitz 1958), the scattering is isotropic if
\begin{equation}
\left(\frac{mc^2}{1\;{\rm GeV}}\right)^{\!3/2}
\left(\frac{u}{2400\;{\rm km}\;{\rm s}^{-1}}\right) 
\left(\frac{\sigma/m}{50\;{\rm cm}^2\,{\rm g}^{-1}}\right)^{\!1/2} \ll 1
\label{eq:slow}
\end{equation}
Note that if the DM particles are WIMPs with the currently favored mass ($m
c^2 >40$ GeV, Hagiwara et al.\ 2002), this condition is not satisfied for
the values of $\sigma/m \sim (1-10)$ \cmg\ that we are able to constrain.
Scattering with such combinations of $\sigma$ and $m$\/ would be strongly
beamed, reducing all the observable effects for a given $\sigma$.  However,
there is still ample parameter space for isotropic scattering (e.g., if
particles are not WIMPs), and we will confine our estimates to this simple
case. Our conclusions will be directly comparable to other astrophysical
cross-section measurements which also implicitly assume isotropic
scattering.

In this reference frame, particle 1 scatters by an angle $\theta=2\alpha_1$
($0<\theta<\pi$) and particle 2 scatters by \mbox{$\pi-\theta$}.  Combining
eqs.\ (\ref{eq:pboth}, \ref{eq:pone}), weighting with solid angle and taking
into account the symmetries, the average momentum lost by the subcluster in
each particle collision is (for our values of $V$ and $v_s$):
\begin{equation}
\bar{p} 
%
= m\vs \left[ 1-4 \int_{\sin\alpha_{\rm e}}^1 
\!\!\!\!x^2 \left\{x^2 -\frac{V^2}{\vs^2}\left(1-x^2\right)\right\}^{1/2}
\!\!dx \right] 
\approx 0.1\,m\vs.
\label{eq:pavg}
\end{equation}

When the subcluster with mass $M_{\rm s}$ (all in the form of dark matter
for clarity) travels with a velocity $v$\/ through the main cluster with
density $\rho_{\rm m}$, it experiences
\begin{equation}
n=\frac{M_{\rm s}}{m}\, \frac{\sigma}{m}\, \rho_{\rm m} v
\end{equation}
collisions per second. As a result, it loses velocity, relative to the
free-fall velocity $v_{\rm ff}$:
\begin{equation}
\frac{d(v-v_{\rm ff})}{dt} = \frac{\bar{p} n}{M_{\rm s}}
= \frac{\bar{p}}{m}\, \frac{\sigma}{m}\, \rho_{\rm m} v,
\label{eq:vloss}
\end{equation}
where $\bar{p}$\/ is from eq.\ (\ref{eq:pavg}).  We conservatively disregard
the higher subcluster velocity during core passage and assume that the loss
of mass and velocity is relatively small, so $\bar{p}$\/ is roughly
constant.  Integration of eq.\ (\ref{eq:vloss}) along the subcluster
trajectory gives the total velocity loss.  Noting that $\int \rho_{\rm
m}\,v\,dt = \int \rho_{\rm m}\,dl$ is the mass column density of the main
cluster along the subcluster's trajectory, $\Sigma_{\rm m}$, we obtain
\begin{equation}
v-v_{\rm ff} = \frac{\bar{p}}{m}\, \frac{\sigma}{m}\, \Sigma_{\rm m}.
\label{eq:deltav}
\end{equation}

Almost all of $\Sigma_{\rm m}$ accumulates within the main cluster's core,
at distances smaller than the current subcluster position.  If, as we
assume, the subcluster passed through the main cluster center, $\Sigma_{\rm
m}\simeq 0.3$ \gcmsq; if it missed the center by $200-300$ kpc, $\Sigma_{\rm
m}$ is lower by about a factor of 2.  On the other hand, we observe the main
cluster after the collision; if it used to have an NFW-type peak disrupted
by a direct hit of the subcluster, then $\Sigma_{\rm m}$ that we should use
would be higher than the observed one.

Requiring that $v-v_{\rm ff} <1000$ \kms, from eq.\ (\ref{eq:deltav}) we get
\begin{equation}
\frac{\sigma}{m} < 7\;{\rm cm}^2\,{\rm g}^{-1}.
\label{eq:cross7}
\end{equation}

\subsection{The survival of the dark matter subcluster}
\label{sec:surviv}

For the observed subcluster mass and velocity, the most likely result of a
particle collision is the loss of a particle by the subcluster.  The
opposite effect, i.e., accretion of the main cluster's particles, is
negligible because of the low mass and high velocity of the subcluster. We
can put an upper limit on the integrated mass loss and thereby on the
collision cross-section.  C04 have derived mass-to-light ratios for the
subcluster within $\rtr=150$ kpc, $M/L_B\simeq 280\pm 90$ and $M/L_I\simeq
170\pm 50$.  These ratios are in good agreement with the universal cluster
values from the lensing analyses (e.g., Mellier 1999; Dahle 2000), and a
factor of $1.1\pm0.3$ from the main cluster values derived from the same
data.  If the subcluster had been continuously losing DM particles, we would
expect an anomalously low $M/L$ value for the subcluster --- and in
particular, a value lower than the main cluster's.  From this agreement we
can infer that the subcluster could not have lost more than $f\approx
0.2-0.3$ of its initial mass within $\rtr$.  We should make two supporting
comments here --- first, the subcluster crossing time for its member
galaxies is comparable to the time it took for the subcluster itself to
cross the main cluster. Thus, if the subcluster's gravitational well has
been slowly diminishing due to DM collisions, its galaxies did not have time
to evaporate, so this process can be ignored.  Second, although the main
cluster's DM particles are similarly affected by the subcluster impact, the
effect on the main cluster's $M/L$ value is smaller by at least their mass
ratio (about a factor of 6).

To escape beyond $r=\rtr$, the average subcluster particle needs a velocity
$v_{\rm esc}\simeq 1200$ \kms.  The subcluster experiences a net loss of a
particle in a collision if both $v_1>v_{\rm esc}$ and $v_2>v_{\rm esc}$,
which occurs at scattering angles $\theta$\/ given by (from eqs.\
\ref{eq:v}):
\begin{equation}
\frac{v_{\rm esc}}{v_0} < \sin\frac{\theta}{2} < 
\left(1- \frac{v_{\rm esc}^2}{v_0^2}\right)^{1/2}.
\label{eq:1-A}
\end{equation}
For our $v_{\rm esc}$ and $v_0$, such angles exist.  The probability of this
loss, per collision, is
\begin{equation}
\chi = \frac{\int_{(16)} 2\pi\,\sin\theta\, d\theta}{
       \int_{\,0}^{\pi} 2\pi\,\sin\theta\, d\theta}
     = 1-2\frac{v_{\rm esc}^2}{v_0^2},
\label{eq:chi}
\end{equation}
where the upper integral is calculated over the interval of $\theta$ given
by eq.\ (\ref{eq:1-A}).  During the subcluster's transit through the main
cluster, each of its original particles has a probability to collide
\begin{equation}
\tau_{\rm m}=\frac{\sigma}{m}\,\Sigma_{\rm m}.
\label{eq:taum}
\end{equation}
Combining (\ref{eq:chi}) and (\ref{eq:taum}), the fraction of particles lost
is
\begin{equation}
\chi \tau_{\rm m} = \frac{\sigma}{m}\,\Sigma_{\rm m}
\left[1-2\left(\frac{v_{\rm esc}'}{v_0}\right)^2\right],
\label{eq:chitau}
\end{equation}
where $v_{\rm esc}'$ is a slightly higher escape velocity that allows for
the subcluster's mass decline by up to a factor of $1+f$ along the way;
$v_{\rm esc}'\approx v_{\rm esc}(1+\sqrt{1+f})/2$.  We also ignore the
subcluster's higher velocity at core passage, assuming the constant present
velocity. (Both effects are small and our simplifications are conservative.)
Requiring that $\chi \tau_{\rm m} <f=0.3$, from eq.\ (\ref{eq:chitau}) we
obtain
\begin{equation}
\frac{\sigma}{m}<1\;{\rm cm}^2\,{\rm g}^{-1}.
\label{eq:cross2}
\end{equation}

The expression in the brackets in eq.\ (\ref{eq:chitau}) is close to 1 for
any small value of $v_{\rm esc}$, thus our resulting constraint is only
weakly dependent on the exact mass profile of the subcluster.  We also note
that the subcluster's $M/L$ ratios derived by C04 represent lower bounds on
the true values (because of possible projection and the particular technique
employed to derive the subcluster mass), which makes our constraint
conservative.

For cross-sections such as (\ref{eq:cross2}) and the subcluster's observed
mass and radius, the mean scattering depth of the subcluster is $\tau_{\rm
s} \lax 0.2$.  Thus our single-scattering assumption is reasonable as a
first approximation.  If one considers a possibility that the escaping
particle may expel another subcluster particle, the probability
(\ref{eq:chitau}) is a conservative underestimate.  In \S\ref{sec:vel}, even
higher $\tau_{\rm s}$ are allowed by our resulting limit (\ref{eq:cross7});
for that method, single scattering is also a conservative assumption.
Inclusion of multiple scattering would lead to tighter limits, but requires
detailed knowledge of the matter distribution inside the subcluster and is
not warranted by the present data.

\section{DISCUSSION}

\subsection{Improving the constraints}

From the above order-of-magnitude estimates, the most promising way of
improving the limit on (or measuring) the DM collision cross-section using
\1e\ would be to refine the weak lensing mass map and limit (or detect) the
mass loss from the subcluster core.  Indeed, if it had lost a significant
fraction of its DM particles, they should form a tail detectable in the dark
matter map, not unlike the X-ray gas tail.  This tail would, of course, be
in addition to the matter tidally stripped from the subcluster and a
gravitational wake discussed by Furlanetto \& Loeb (2002).  The main
cluster's DM particles also would be scattered, probably resulting in
additional detectable effects in the mass map.  These considerations, along
with the conservative assumptions that we made throughout to simplify
calculations, suggest that the best way to proceed with the data
interpretation is a detailed hydrodynamic simulation of this merger, such as
that performed by Tormen, Moscardini, \& Yoshida (2003) but with the
inclusion of collisionless galaxies and collisional DM with various
cross-sections, followed by comparison of the results with the X-ray,
optical and lensing data.

\subsection{Astrophysical context}
\label{sec:disc}

The nature of dark matter is still a great unsolved astrophysical problem.
Laboratory searches for the DM candidates succeeded in showing that any
interaction between the dark and baryonic matter is vanishingly small, with
cross-sections many orders of magnitude lower than the values we consider in
this work (e.g., Bernabei et al.\ 2003 and references therein).  Despite
their size, galaxy clusters have an average projected mass density of order
$0.1-1$ \gcmsq, so they are no match for the laboratory experiments for
constraining the DM-baryon interactions.  However, clusters and galaxies may
provide the best available laboratory for studying the DM self-interaction.

While the common wisdom holds that DM is collisionless, a hypothesis of
self-interacting dark matter (SIDM) with cross-sections of order $1-100$
\cmg\ was most recently proposed by Spergel \& Steinhardt (2000) to
alleviate several apparent problems of the collisionless CDM model, such as
the non-observation of predicted cuspy mass profiles in galaxies (e.g.,
Moore 1994; Flores \& Primack 1994; cf.\ Navarro, Frenk, \& White 1997;
Moore et al.\ 1999b) and over-prediction of small sub-halos within the
larger systems (e.g., Klypin et al.\ 1999; Moore et al.\ 1999a).
Simulations and theoretical studies (e.g., Dav\'e et al.\ 2001; Ahn \&
Shapiro 2002) narrowed the range required to explain the galaxy profiles to
$\sigma/m \sim 0.5-5$ \cmg\ and pointed out that fluid-like DM with
$\sigma/m \sim 10^4$ \cmg\ is another possibility.

Several observational constraints have been reported.  Gnedin \& Ostriker
(2001; see also Hennawi \& Ostriker 2002) pointed out that unless
$\sigma/m<0.3-1$ \cmg, galactic halos inside clusters should evaporate on
the Hubble timescale, because of collisional heat conduction from the hot
cluster DM particles into the cool halos.  We note that our constraint in
\S\ref{sec:surviv} uses a very similar idea, except that we have a
cluster-sized halo subjected to a flow of particles from one direction
instead of random bombardment.

Furlanetto \& Loeb (2002) proposed to use the different shapes of a
gravitational wake that a subcluster halo moving through a bigger halo would
create in the collisionless and fluid-like DM models.  They argue that the
X-ray image of the bright galaxy in the Fornax cluster already disfavors
fluid-like DM.  Their method is somewhat similar to our \1e\ estimates in
that it also uses the motion of a halo within a bigger halo, but appears
more observationally challenging.  Following another suggestion of
Furlanetto \& Loeb (2002), Natarajan et al.\ (2002) used observed sizes of
the galactic halos in the A2218 cluster (which should be truncated at
different radii for different DM cross-sections) to obtain $\sigma/m<40$
\cmg.  Upper limits in the $0.02-10$ \cmg\ range were derived by Hennawi \&
Ostriker (2002) from the absence of supermassive black holes in the centers
of galaxies.

Strong constraints are reported from the galaxy cluster cores.  Yoshida et
al.\ (2000) simulated evolution of a cluster for $\sigma/m=10$, $1$, and
$0.1$ \cmg\ and obtained systematically flatter radial mass profiles for
higher $\sigma/m$.  Using \chandra\ X-ray data and the assumption of
hydrostatic equilibrium, Arabadjis et al.\ (2002) derived a mass profile for
the cluster MS\,1358+62 that is strongly centrally peaked (in fact, more
peaked than even the collisionless simulations predict). From the comparison
with Yoshida et al.\ (2000), they conclude that $\sigma/m<0.1$ \cmg. There
are two potential difficulties with this stringent limit, however.
\chandra\ revealed widespread gas sloshing in the cores of ``relaxed''
clusters (Markevitch, Vikhlinin, \& Forman 2002), including in MS\,1358+62,
which does not lend support to the hydrostatic equilibrium assumption and
the resulting X-ray mass estimates at relevant radii.  More importantly,
Yoshida et al.\ present time evolution of their simulated cluster profile
for $\sigma/m=10$ \cmg, which at different epochs covers a range of shapes
comparable to the whole difference between their 0.1 and 10 \cmg\
simulations (see also Hennawi \& Ostriker 2002).  This suggests that for a
high $\sigma/m$, one expects a variety of cluster mass profiles --- and
indeed, while many real-world clusters exhibit peaked profiles, a large
fraction has flat cores (e.g., Coma). Therefore, a sample of cluster
profiles is required for such studies.  Miralda-Escud\'e (2002) used lensing
data for the cluster MS\,2137--23 to show that its central mass distribution
is elliptical, and concluded that $\sigma/m<0.02$ \cmg, since otherwise DM
collisions would erase ellipticity.  Again, this interpretation may not be
unique, because ellipticity may be affected by line-of-sight projections.
This ambiguity is illustrated by the fact that Sand et al.\ (2002), using
lensing data for the same cluster but considering the flatness of the
central mass distribution instead of its ellipticity, arrived at the
opposite conclusion.

While the above methods may eventually provide more sensitive constraints on
$\sigma/m$ than we have obtained, they require a statistical sample of
clusters and cosmological simulations to reach solid conclusions.  In
comparison, the unique geometry of the \1e\ cluster merger allows us to see
directly the (cumulative) results of single DM particle collisions, which
makes the resulting limits quite robust.

Finally, we note that our limit, $\sigma/m<1$ \cmg, excludes most of the
$0.5-5$ \cmg\ interval proposed to explain the flat mass profiles in
galaxies.  Within the SIDM paradigm, the galaxy profiles and the tight
cross-section limits coming from clusters can still be reconciled if the
cross-section were velocity-dependent, so that it would be smaller on
average in clusters than in galaxies (e.g., Firmani et al.\ 2000, 2001;
Hennawi \& Ostriker 2002; Col{\'{\i}}n et al.\ 2002).  However, it is
difficult to justify this additional degree of freedom in the model until a
nonzero cross-section is detected at any velocity.

\section{SUMMARY}

We have combined new X-ray, optical and weak lensing observations of the
unique merging cluster \1e\ to derive a simple, direct upper limit on the
dark matter collisional cross-section, $\sigma/m<1$ \cmg.  This is only an
order-of-magnitude estimate; a more accurate, and quite possibly stronger,
limit may be derived through hydrodynamic simulations of this merging
system.

\acknowledgements

We thank Neal Dalal and the anonymous referee for helpful comments.  Support
for this work was provided by NASA contract NAS8-39073, \chandra\ grant
GO2-3165X, and the Smithsonian Institution.  AHG was supported under award
AST-0407485 by an NSF Astronomy and Astrophysics Postdoctoral Fellowship; DC
received support from Deutsche Forschungsgemeinschaft under the project SCHN
342/3-1.

\end{document}